\documentclass[10pt,journal, a4paper]{IEEEtran}
\usepackage{amssymb,amsmath,bm}
\usepackage{amsfonts}
\usepackage{cite}
\usepackage{graphicx,subfigure}
\usepackage{psfrag}
\usepackage{url}
\usepackage[latin1]{inputenc}
\usepackage[absolute,overlay]{textpos}
\usepackage{tikz}
\usetikzlibrary{arrows,calc,decorations.markings}
\usetikzlibrary{topaths}
\usetikzlibrary{arrows}
\usetikzlibrary{shadows}
\usetikzlibrary{positioning}
\usetikzlibrary{matrix}
\usetikzlibrary{shapes.geometric}
\usetikzlibrary{decorations.pathmorphing}
\usetikzlibrary{calc}
\usetikzlibrary{fit}					
\usetikzlibrary{backgrounds}	

\usepackage{pgf}
\usetikzlibrary{arrows,automata}
\usepackage[latin1]{inputenc}
\usepackage[nocomma]{optidef}

\usepackage{algorithm}
\usepackage{algorithmic}

\makeatletter
\def\thickhline{%
  \noalign{\ifnum0=`}\fi\hrule \@height \thickarrayrulewidth \futurelet
   \reserved@a\@xthickhline}
\def\@xthickhline{\ifx\reserved@a\thickhline
               \vskip\doublerulesep
               \vskip-\thickarrayrulewidth
             \fi
      \ifnum0=`{\fi}}
\makeatother

\newlength{\thickarrayrulewidth}
\usepackage{threeparttable}
\usepackage{booktabs}

\usepackage{arydshln}

\usepackage{amsthm}

\begin{document}

\title{On the Optimal Deployment of Power Beacons for Massive Wireless Energy Transfer}

\author{Osmel Mart\'{i}nez Rosabal,
        Onel L. Alcaraz L\'{o}pez, Hirley Alves, Samuel Montejo-S\'{a}nchez, Matti Latva-aho
\thanks{Osmel Mart\'{i}nez Rosabal,
        Onel L. Alcaraz L\'{o}pez, Hirley Alves, and Matti Latva-aho are with the Centre of wireless Communications (CWC), University of Oulu, Finland. \{osmel.martinezrosabal, onel.alcarazlopez, hirley.alves, matti.latva-aho\}@oulu.fi}
\thanks{Samuel Montejo-S\'{a}nchez is with Programa Institucional de Fomento a la I+D+i, Universidad Tecnol\'{o}gica  Metropolitana, Santiago 8940577, Chile. smontejo@utem.cl}
\thanks{This research has been financially supported by Academy of Finland, 6Genesis Flagship (Grant n.318927), EE-IoT (n.319008), Aka Prof (n.307492), as well as FONDECYT Iniciaci\'{o}n No. 11200659, FONDECYT Regular No. 1201893, and  FONDEQUIP EQM180180.}}

\maketitle

\begin{abstract}
Wireless energy transfer (WET) is emerging as an enabling green technology for Internet of Things (IoT) networks. WET allows the IoT devices to wirelessly recharge their batteries with energy from external sources such as dedicated radio frequency transmitters called power beacons (PBs). In this paper, we investigate the optimal deployment of PBs that guarantees a network-wide energy outage constraint. Optimal positions for the PBs are determined by maximizing the average incident power for the worst location in the service area since no information about the sensor deployment is provided. Such network planning guarantees the fairest harvesting performance for all the IoT devices. Numerical simulations evidence that our proposed optimization framework improves the energy supply reliability compared to benchmark schemes. Additionally, we show that although both, the number of deployed PBs and the number of antennas per PB, introduce performance improvements, the former has a dominant role. Finally, our proposal allows to extend the coverage area while keeping the total power budget fixed, which additionally reduces the level of electromagnetic radiation in the vicinity of PBs.
\end{abstract}
\section{Introduction}
The broad range of today's Internet of Things (IoT) applications demands a massive deployment of low-cost devices, powered by small batteries, and with different quality-of-service (QoS) requirements. Battery lifetime is critical, since maintenance operations could be impractical, if not forbidden, either in massive or harsh conditions deployments \cite{8030485}. In order to extend the battery lifetime, or even avoid completely its need, energy harvesting techniques, from renewable sources such as solar and wind, or from ubiquitous radio-frequency (RF) signals, have been recently proposed and investigated \cite{8421584}. Dedicated RF sources provide an alternative for supplying controllable and reliable energy for low-power IoT devices. Wireless energy transfer (WET) techniques are therefore important enablers of many coming IoT use cases \cite{lopez2019massive}. Although there are many WET techniques, herein we focus exclusively on RF WET, and hereafter WET refers to RF-based energy transfer.

In \cite{7164906}, the authors consider a scenario where PBs transfer energy to rechargeable devices considering practical technology limitations. Therein, they were interested in maximizing the energy transferred while constraining the maximum level of radio-frequency-electromagnetic field (RF-EMF) due to its risk to human health. A similar concern is exposed in \cite{8485951}, where the authors maximize the charging utility function of the energy harvesters (EHs) while constraining the level of RF-EMF through a meta-distribution characterization of the per-sensor received power. 

Nature-inspired algorithms have been recently proposed in \cite{9165837} for solving the placement problem of PBs. Therein, the authors propose a WET solution for extending the lifetime of multihop networks. They propose an algorithm that mimics the flocking behavior in order to maximize the network lifetime by equalizing the lifetime of individual sensors. Moreover, the use of multiple antennas provides diversity over the space as well as additional degrees of freedom, suitable to tackle the impairments caused by small-scale fading. This applies to wireless powered communication networks (WPCNs), where beamforming techniques can be implemented to increase the coverage area due to the directionality in transmission \cite{8269301}. In \cite{6623072}, it is considered an energy beamforming technique to tackle the problem of long-distance power transfer in large-scale multiple-input multiple-output (MIMO) systems. In \cite{8444982}, a wirelessly powered massive MIMO system, where a multi antenna base station (BS) charges single-antenna EHs in the downlink, is optimized. A distributed wireless power transfer system is studied in \cite{8249549} and \cite{7993879} where the focus is optimal distributed energy beamforming to overcome the effects of destructive interference whether frequency and phase synchronization is available or not. They conclude that distributed antenna systems are more effective than collocated antennas at a single PB as long as the optimal inter-PB beamforming can be achieved. To improve the energy transfer efficiency in a large WPCN, \cite{7417795} proposes an adaptative energy beamforming strategy that maximizes the average received power by exploiting the tradeoff between the power intensity of the beams and the number of sensors to be charged. In \cite{8930937}, the authors propose a charge scheduling scheme using battery-powered PBs which harvest energy from a central PB. They exploit the benefits of beamforming in order to maximize the energy efficiency while satisfying QoS constraint of the sensor network. A max-min rate problem is addressed in \cite{9099249}, where solar-powered multi-antenna PBs assist an IoT network. Therein, the authors propose a distributed rate allocation protocol that considers the state of the data buffer and battery level information of each device and its nearest neighbors, in order to forward the collected measurements.

However, the above solutions require perfect channel state information (CSI) acquisition for estimating the optimal beamforming vector, which consumes more energy as the number of EHs increases \cite{6994766}. In an attempt to fill this gap, the authors of \cite{8760520},\cite{9119347} study a WET setup where a multi-antenna PB operates without CSI for powering massive low-power IoT deployments. They analyze the statistics of the harvested energy under the sensitivity and saturation phenomena, and compare it with that experienced when using CSI-based schemes. Particularly, in \cite{9119347} the authors do consider the phase shift between antenna elements and compare it with the assumption of independent transmitted signals.

Hybrid networks deployments under outage constraints are investigated in \cite{6697937}, where authors assume that single-antenna BSs and PBs form independent homogeneous Poisson point processes (PPP). Interesting tradeoffs between transmit power and density of BSs and PBs are derived under an outage constraint on the data link. Multiple antennas at the BS and PBs are considered in \cite{8645821}, where the latter are deployed following a PPP with certain density. 

Therein, the authors investigate the area that should be covered by the centered BS to minimize the average power consumption, so that the uplink rate is above certain threshold with probability given by the QoS requirement of all devices in the network.
\begin{table*}[t]
    \centering
    \caption{Related Works and Main Modeling Assumptions.}
    \label{tab:works}
    \begin{tabular}{l l l l l}
        \thickhline
            \textbf{Reference} & \textbf{Sensors' positions} & \textbf{Antennas} & \textbf{PBs Mobility} & \textbf{Objective} \\
        \hline
            \cite{8428435} & Known & Directional & Static & maximize the average harvested energy \\
            \cite{8714083} & Known & Directional & Static & maximize the weighted-sum of the network-wide harvested energy \\
            \cite{8328826} & Known & Onmidirectional & Static & minimize the number of PBs that guarantees full coverage \\
            \cite{9155356} & Known & Directional & Mobile & maximize the network-wide harvested energy \\
            \cite{bi2015placement} & Known & Onmidirectional & Static & minimize the cost of network deployment \\
            \cite{9209664} & Not necessarily known & Directional & Mobile & minimize the charging time \\
            Current work & Unknown & Onmidirectional & Static & minimize the number of PBs that satisfies an energy outage constraint \\
        \thickhline
    \end{tabular}
\end{table*}
\subsection{Related works}
Recently, PBs deployment optimization emerged as a potential technique to ban blind-spots and homogenize the network energy availability. For instance, in \cite{8428435} the authors take into account directional antenna patterns to optimize the placement of PBs in order to maximize a charge utility function for all deployed sensors. Therein, they consider the benefits of directivity over the omnidirectional radiation, and propose an algorithm to dynamically place the PBs as the network topology changes. Similarly, the authors in \cite{8714083} optimize the positions of PBs considering the presence of obstacles. Therein, they maximize the network-wide weighted sum harvested energy with heterogeneous PBs and sensors.

In order to enhance coverage, Daubechies wavelet algorithm is proposed in \cite{8328826} to minimize the number of PBs to be deployed. Assuming the sensors' deployment to be known, they find the optimum PBs' positions by moving the redundant PBs at each iteration towards the uncovered sensors. Meanwhile, the authors in \cite{9155356} study the problem of mobile directional PBs' deployment to improve energy coverage. Each PB can move inside certain area such that the overall harvested energy is maximized. In \cite{bi2015placement}, the authors optimize the number and locations of energy and information access points in a WPCN subject to energy harvesting and communication performance requirements. The authors associate a cost to each node's installation to then minimize the overall network cost deployment. They address the network performance given average channel gains and considering wireless devices' positions to be known a priori, which allows to divide the network into non-overlapping clusters. Hence, the proposed algorithms yield different solutions depending on the sensors' locations for the same network area, and therefore it could not be optimal if these locations are not predetermined. Finally, the position and speed of a single mobile PB powering a sensors' deployment is optimized in \cite{9209664}. In this case, authors don't rely on the sensors' positions. Instead, they divide the service area into smaller sub-partitions and focus on the performance of the instantaneous worst position, which changes as the PB moves. The goal is to minimize the overall charging time constraining the movement of the PB.

\subsection{Motivation and contributions of this work}
Different from above works, herein, we study the optimal deployment of PBs to satisfy an energy outage constraint. We consider a massive deployment of EHs for which CSI acquisition procedures are too costly, therefore, instantaneous CSI is not available. Since neither CSI nor devices locations are known, PBs' positions are optimized for maximizing the foreseen minimum average RF energy available in the area. This guarantees for each EH to most likely meet its corresponding QoS requirements. Moreover, our proposed strategy does not depend on neither the hardware heterogeneity nor the mobility of the sensors. Table~\ref{tab:works} lists the main differences of our model compared with previous related works.

Our main contributions are summarized below
\begin{itemize}
    \item We provide a framework for assessing the performance of a WET setup operating with the minimum number of PBs required to support energy QoS requirements;
	\item We discuss and evaluate several methods for finding the optimal deployment of PBs for powering a massive number of sensors;
	\item Our results suggest that in case of one and two PBs, both should be deployed in the center of the $R$ radius area under evaluation, while in case of three and four PBs, they should be symmetrically located in a concentric circumference approximately of radius $\frac{R}{2}$ and $\frac{R \sqrt{2}}{2}$, respectively. For a greater number of PBs, their optimal positions depend strictly on the area and path loss exponent;	
	\item Numerical results show that the number of PBs deployed in the network has a greater impact on the system performance than the number of antennas per PB, even though both introduce improvements in the energy outage.
\end{itemize}

\subsection{Organization of the paper}
Next, Section \ref{system} introduces the system model and presents the problem formulation. In Section \ref{PBoptimum}, we discuss several approaches for finding the optimal positions for a fixed set of PBs, while in Section \ref{numerical} we show and analyze numerical performance results as a function of the number of PBs and available transmit antennas. Finally, Section \ref{conclusions} concludes the paper.

\textit{Notation:} Here, we use boldface lowercase letters to denote column vectors. The operator $|\cdot|$ can represent either the absolute value for scalars or the cardinality of a set, while $\lVert x \lVert_p = (\sum_{\forall i \ge 1} \vert x^p_i \rvert)^{1/p}$ denotes the $\mathcal{\ell}_p$-norm \cite[eq. (1)]{lange2013derivatives}. Given a minimization problem with objective function $f_0(x)$, $m$ inequality constraints $f_i(x) \leq 0, \ i = 1, \ldots, m$, and $n$ equality constraints $h_i(x), \ i = 1, \ldots, n$, we denote $\mathcal{L}(x, \bm{\nu}, \bm{\lambda}) = f_0(x) + \sum_{i=1}^{m} \nu_i f_i + \sum_{i=1}^{n} \lambda_i h_i(x)$ as the Lagrangian of the problem \cite{boyd2004convex}. Herein, $\nu_i$ and $\lambda_i$ are the Lagrange multipliers of the $i^\text{th}$ inequality and equality constraints, respectively. The symbol $\nabla$ represents the gradient operator, i.e. $\nabla f(x_0):\mathbb{R}^n \rightarrow \mathbb{R}^n$ is the gradient of the function $f(x):\mathbb{R}^n \rightarrow \mathbb{R}$ at the point $x_0$. Moreover, $\mathcal{O}(\cdot)$ is the big-O notation, which specifies worst-case complexity. Finally, $z\!\sim\!\mathcal{N}(\mu, \sigma^2)$ is a Gaussian distributed random variable with 
mean $\mathbb{E}[z] = \bm{\mu}$ and variance $\sigma^2$, while $y \sim \chi^2(k, \phi)$ is a non-central chi-squared random variable with $k$ degrees of freedom, and noncentrality parameter $\phi$ \cite{krishnamoorthy2016handbook}. Table.~\ref{tab:sym} lists the symbols used throughout the paper.
\begin{table}[t]
    \centering
    \caption{List of symbols}
    \label{tab:sym}
    \begin{tabular}{p{1.2cm} p{7cm}}
        \thickhline
            \textbf{Parameter} & \textbf{Definition} \\
        \hline
            $\mathrm{EH}_s$ & $s^\text{th}$ EH in the set $\mathcal{S}$ \\
            $\mathrm{PB}_b$ & $b^\text{th}$ PB in the set $\mathcal{B}$ \\
            $h_{s,b}$ & complex channel coefficient of the $\mathrm{PB}_b \rightarrow \mathrm{EH}_s$ link \\
            $\alpha$, $\beta$ & real and imaginary parts of $h_{s,b}$ respectively \\
            $\varrho_{s,b}$ & path-loss function of the $\mathrm{PB}_b \rightarrow \mathrm{EH}_s$ link\\
            $d_{s,b}$ & distance of the $\mathrm{PB}_b \rightarrow \mathrm{EH}_s$ link\\
            $P$, $P_T$ & PB transmit power and total transmit power respectively \\
            $\kappa$ & Rician fading factor \\
            $\gamma$ & path-loss exponent \\
            $K$ & unitless constant for the path-loss model \\
            $x_{b}$ & energy-carrying signal transmitted from $\mathrm{PB}_b$ \\
            $\mathcal{A}$ & set of antennas at each PB \\
            $\xi^\mathrm{RF}_s$ & incident RF power at the $s^\text{th}$ sensor \\
            $\xi_0$ & sensitivity of the EH's circuitry \\
            $\zeta$ & energy outage probability threshold \\
            $R$ & radius of the coverage area \\
            $T_\mathrm{avg}$ & normalized average computational time \\
            $\bm{\lambda}$ & Lagrange multiplier vector for equality constraints, $\bm{\lambda} = \{\lambda_b\}$ \\
            $\mathbf{t}$ & slack variable, $\mathbf{t} = \{t_b\}$ \\
            $\mu$ & barrier parameter \\
        \thickhline
    \end{tabular}
\end{table} 
\section{System model}\label{system} 
\begin{figure}[t]
	\centering
	\includegraphics[width=0.4\textwidth]{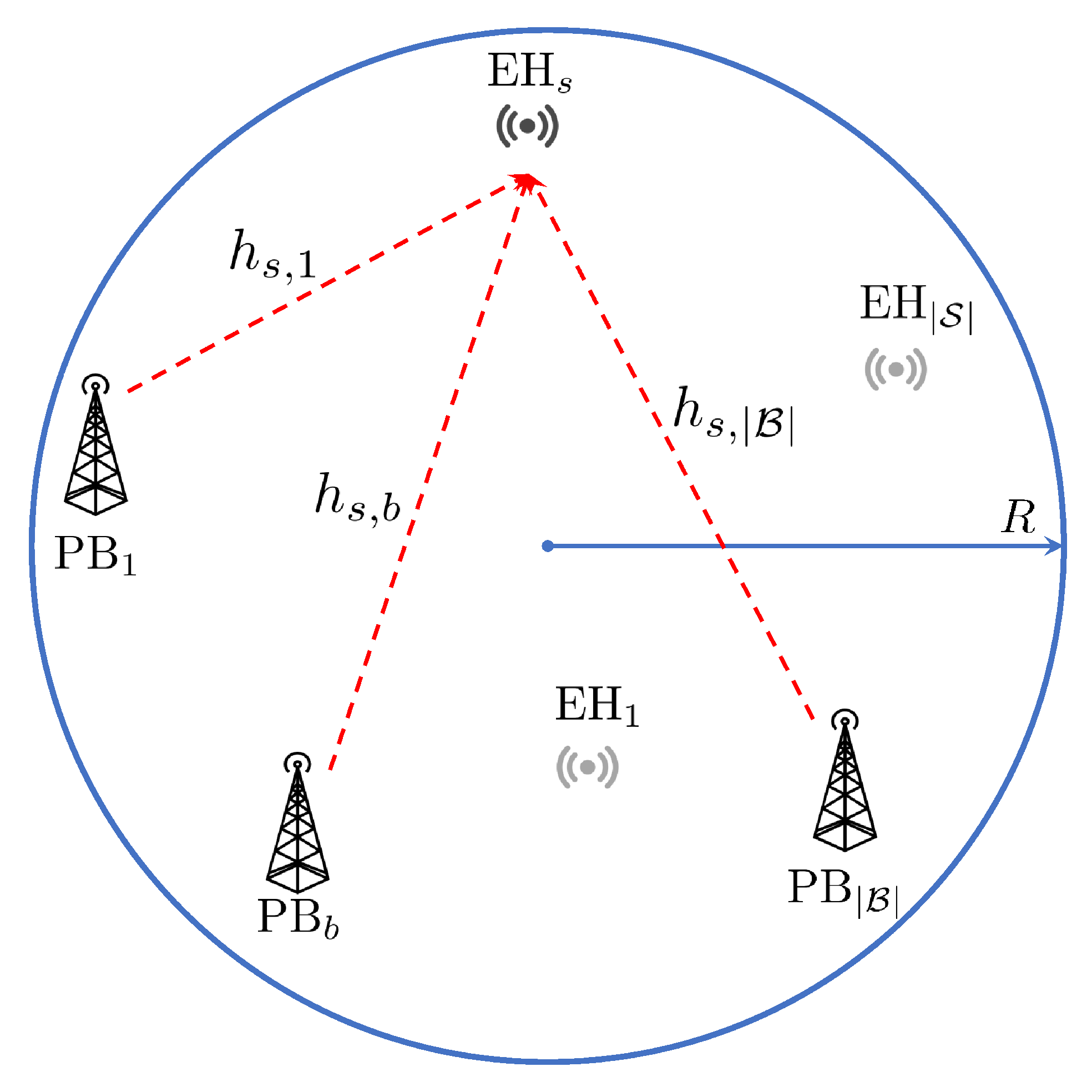}
	\vspace*{-4mm}
	\caption{The system model comprises a set $\cal B$ of PBs deployed to charge wirelessly a massive set ${\cal S}$ of IoT energy harvesters.}
	\vspace*{-2mm}		
	\label{scenario}
\end{figure}

Consider the scenario in Fig.~\ref{scenario} where a set $\mathcal{B} = \{\mathrm{PB}_b|b = 1,2,\ldots,|\cal B| \}$ of PBs are deployed in a circular area of radius $R$ to wirelessly power a massive number of EHs, denoted by a set $\mathcal{S} = \{\mathrm{EH}_s|s=1,2,\dots,|\mathcal{S}| \}$. Each PB is equipped with an omnidirectional antenna, and transmits at the same power level $P$. We also assume quasi-static channels with independent and identically distributed (i.i.d) Rician fading with factor $\kappa$. The channel coefficient between $\mathrm{EH}_s \in \cal S$ and the $\mathrm{PB}_b \in \cal B$ is denoted as $h_{s,b} \in \mathbb{C}$, and $h_{s,b} = \alpha + j \beta$ with independent real and imaginary parts $\alpha, \beta \sim \mathcal{N}\Big(\sqrt{\frac{\kappa}{2 (1 + \kappa)}}, \frac{1}{2 (1 + \kappa)}\Big)$ \cite{7225184}. There is no knowledge about sensors' positions, and PBs are not able to estimate the CSI due to the large number of devices and/or their stringent associated energy expenditure constraints \cite{8760520},\cite{9119347}. We denote the energy-carrying signal coming from the $\mathrm{PB}_b$ as $x_b$ and we consider that independent signals are transmitted from all the sources. Then, the incident RF power at the $s^\text{th}$ harvester is
\begin{equation}
    \xi^\mathrm{RF}_s = \sum_{b=1}^{|\cal B|} \varrho_{s,b} |h_{s,b}|^2 \mathbb{E}[|x_b|^2] = P \sum_{b=1}^{|\cal B|} \varrho_{s,b} |h_{s,b}|^2,
\end{equation}    
where $\varrho_{s,b} = K d_{s,b}^{-\gamma}$ comprises the path-loss of the $\mathrm{EH}_s \rightarrow \mathrm{PB}_b$ link with distance $d_{s,b}$, which depends on the path-loss exponent $\gamma$ and other factors considered in $K$ such as the carrier frequency and antenna gains \cite{goldsmith2005wireless}. Finally, $x_{b}$ is such that  $\mathbb{E} [|x_b|^2] = P, \forall b \in \cal B$, while $P$ is normalized in the time domain, which allows us to indistinctly use the terms energy and power for the same quantity. 
Authors in \cite[eq.(24)]{8760520} found that the distribution of the harvested energy, when a single-antenna PB serves the network, is a scaled non-central chi-squared with non-centrality parameter $2 \kappa$, and $2$ degrees of freedom. We can extend those results to our model and write
\begin{equation}\label{gCHI}
    \xi^\mathrm{RF}_s ~ \sim \sum_{b=1}^{|\mathcal{B}|} \frac{\varrho_{s,b}}{2(1+\kappa)} \chi^2(2, 2\kappa),
\end{equation}
which is a linear combination of i.i.d non-central chi-squared random variables. The sum \eqref{gCHI} is a particular way of defining the generalized chi-square distribution, whose probability density function is often computed by numerical algorithms due to its analytical intractability \cite{ha2013accurate}.

\subsection{Problem formulation}
We consider a massive deployment of EHs in the area. The main goal is to estimate the minimum number of PBs that need to be deployed in order to satisfy an energy outage probability constraint with threshold $\zeta$. The ultimate incident RF energy $\xi^{RF}_s$ must be above the sensors' sensitivity $\xi_0$ with probability $1-\zeta$, $\forall s \in \cal S$. The optimization problem can be formulated as follows
\begin{subequations}\label{P1}
    \begin{alignat}{2}
    \mathbf{P1:} \quad &\mathrm{minimize} \quad && |\mathcal{B}| \label{P1a}\\
    &\text{subject to} \quad &&\mathbb{P}(\xi^{RF}_s \le \xi_0) \le \zeta , \quad \forall s \in \cal S, \label{P1b}\\
    & \quad && P = \frac{P_T}{|\mathcal{B}|}\label{P1c}.
    \end{alignat}
\end{subequations}
Since the domain of the objective is $|\mathcal{B}| \in \mathbb{N}$, \textbf{P1} is a mixed-integer programming problem which is in general nondeterministic polynomial (NP) complete \cite{7742722}. However, we can overcome this issue by searching over the space of candidate solutions for which the positions of the PBs are optimized. Additionally, the constraint \eqref{P1b} is by far cumbersome, so we rely on Monte Carlo simulations when computing the energy outage probability. Finally, \eqref{P1c} refers to the per-PB maximum transmit power given the network power constraint $P_T$.

\section{Optimal PBs positioning}\label{PBoptimum}
The problem \textbf{P1} relies on the optimal positions of PBs given certain propagation conditions. Thus, let us fix $|\cal B|$ to first find the optimal PBs positions. Without loss of generality, we discretize the circular region where sensors are massively located, hence, mimicking its deployment. This is done by creating evenly spaced circumferences with a discrete number of points proportional to its radius.
Let $\mathbf{u}_s, \mathbf{n}_b \in \mathbb{R}^{2 \times 1}$ denote coordinate vectors for the locations of $\mathrm{EH}_s$ and $\mathrm{PB}_b$, respectively. Notice that as $| \cal S|$ increases, the points' deployment tends to cover the whole area. We can state the average RF energy available at the $s^\text{th}$ sensor, as a function of the distance to each PB as
\begin{align}\label{objFunc}
\mathbb{E}[\xi^\mathrm{RF}_s] = \frac{P_T}{|\mathcal{B}|} K \sum_{b=1}^{|\cal B|} \lVert \mathbf{u}_s - \mathbf{n}_b\rVert^{-\gamma}_2 .
\end{align}
Notice that, in order to meet the system constraints, it is sufficient that the EH performing the worst, e.g., the one under the greatest path loss, meets the energy outage requirement. Then, the PBs positioning optimization problem can be stated as
\begin{subequations}\label{P2}
    \begin{alignat}{4}
    \mathbf{P2:} \quad & \underset{\mathbf{n}_b}{\operatorname{arg\,max}} &&\quad \underset{s}{\mathrm{min}}  \quad \mathbb{E}[\xi^\mathrm{RF}_s],\quad &&&\forall s \in \cal S, \label{P2a} \\
    &\text{subject to} &&\quad \lVert \mathbf{n}_b \rVert_2 - R \leq 0,\quad &&&\forall b \in \cal B, \label{P2b}
    \end{alignat}
\end{subequations}
which maximizes the average received energy at the worst sensor through the optimal deployment of PBs within the coverage area.
\subsection{Equally-far-from-center approach (EC)}\label{EC}
\begin{figure}[t!]
	\centering
	\includegraphics[width=0.4\textwidth]{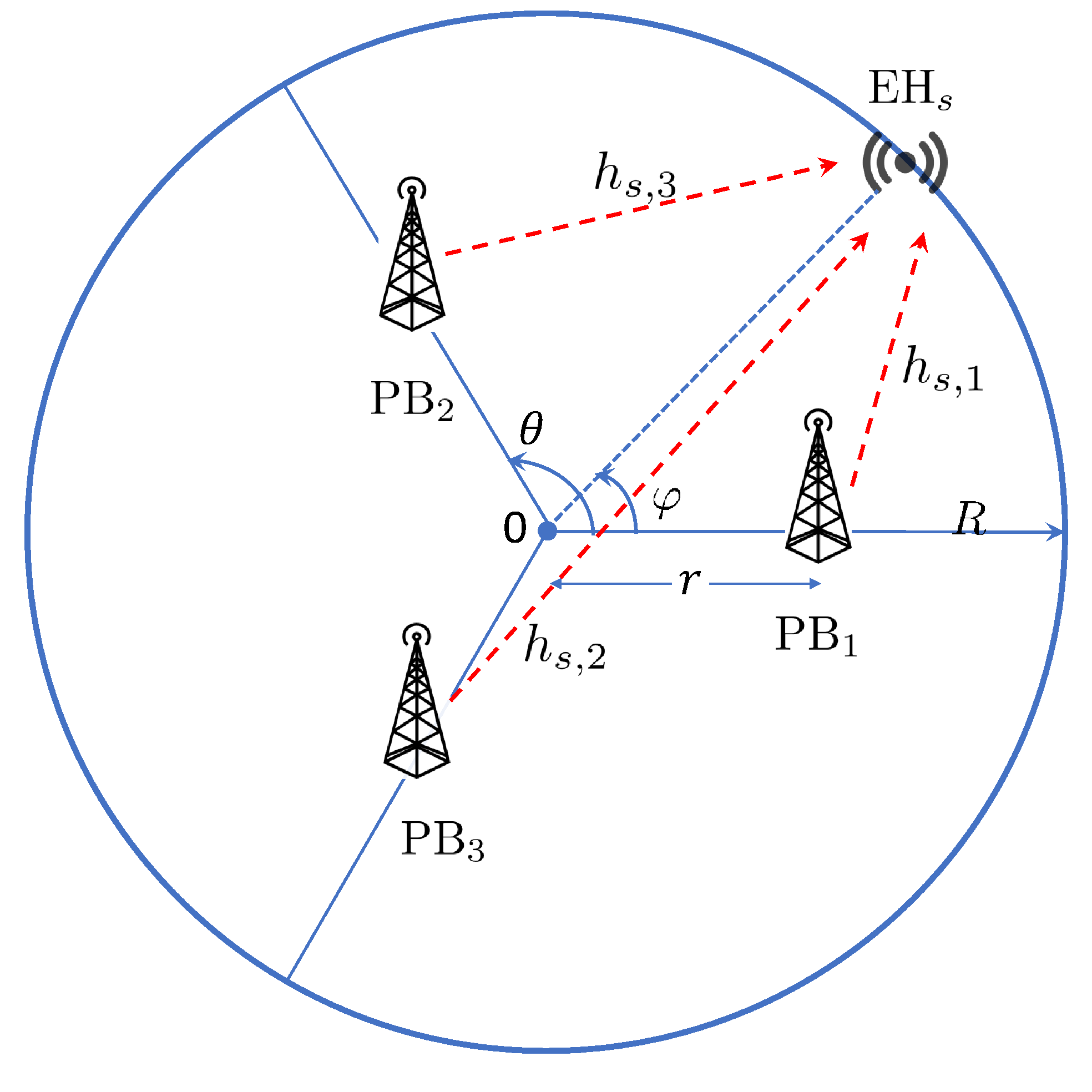}
	\vspace{-3mm}
	\caption{Example of an EC deployment for $|\mathcal{B}| = 3$.}
	\vspace*{-2mm}		
	\label{auxFig}
\end{figure}

We can take advantage of region's symmetry properties to reduce the complexity of the problem. Herein, we assume the PBs to be equally-far-from the circle center at a distance $\lVert \mathbf{n}_{b} \rVert_2 = r: 0 \le r\le R, \forall b \in \cal B$. Note that the angle $\theta$ determined by any two adjacent PBs and vertex at the circle center must obey $\theta = \arccos \frac{\mathbf{n}_b \cdot \mathbf{n}_{b'}}{\lVert \mathbf{n}_b \rVert_2 \cdot \lVert \mathbf{n}_{b'} \rVert_2} = \frac{2\pi}{|\mathcal{B}|}, \forall b,b' \in \cal B$, with $b' \neq b$. Under these conditions, the average receive power is the minimum either at the circle edge or at the circle center. The deployment is illustrated in Fig.~\ref{auxFig} for $|\mathcal{B}| = 3$.
The average received power at the $s^\text{th}$ sensor on the region's edge using polar coordinates is

\begin{small}
    \begin{equation}\label{twoPBs}
        \mathbb{E}[\xi^\mathrm{RF}_{s}] = P K \sum_{b=1}^{|\mathcal{B}|}\Big[ r^2 \!+\! R^2 \!-\! 2rR \cos{\big(\theta(b\!-\!1) \!-\! \varphi \big)} \Big]^{-\frac{\gamma}{2}},
    \end{equation}
\end{small}

\begin{figure}[t!]
	\centering
	\includegraphics[width=0.45\textwidth]{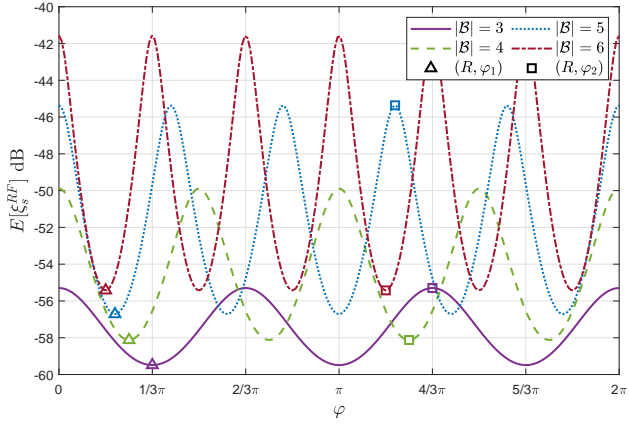}
	\vspace{-3mm}
	\caption{Average incident RF power at the edge vs $\varphi$ for $|\mathcal{B}| \in \{3, 4, 5,6\}$.}
	\vspace*{-2mm}		
	\label{RFatEDGE}
\end{figure}

\hspace{-3.5mm}where $\varphi$ is the angle of the sensor located on the edge. Since the set of signals $\{x_{b} \}$ are independent, we can take advantage of the superposition principle considering first the contribution of $\mathrm{PB}_1$ and $\mathrm{PB}_2$ in $\mathrm{EH}_s$ to find the worst position on the edge. We find the critical points by taking $\frac{\partial \mathbb{E}[\xi^{RF}_{s}]}{\partial \varphi} = 0$ considering the contribution of $\mathrm{PB}_1$ and $\mathrm{PB}_2$ in \eqref{twoPBs}, and after some algebraic manipulations, we attain 
\begin{align}\label{derivativeE}
 \frac{M^{\frac{\gamma}{2} + 1}\sin{\varphi}- N^{\frac{\gamma}{2}+1}\sin{(\theta - \varphi)}}{M^{\frac{\gamma}{2} + 1} N^{\frac{\gamma}{2} + 1}} = 0 ,
\end{align}
where
\begin{align*}
    M &= r^2 + R^2 -2rR\cos{\big(\theta -\varphi \big)},\\
    N &= r^2 + R^2 -2rR\cos{\varphi}.
\end{align*}
The roots of \eqref{derivativeE} are determined by setting its numerator to $0$, which has a trivial solution at $\varphi_{1,2} = \{ \theta/2, \theta/2 + \pi \}$ within $[0, 2\pi]$ when
\begin{equation}
    \Big[\frac{M}{N} \Big]^{-\frac{\gamma}{2}-1} = \frac{\sin{\varphi}}{\sin{\big( \theta - \varphi \big)}},
\end{equation}
which happens when $\frac{M}{N}$ and $\frac{\sin{\varphi}}{\sin{\big( \theta - \varphi \big)}}$ are both equal to one. This result is expected, since it refers to the most distant sensors in terms of symmetric power contribution considering two adjacent PBs. In fact, the path gain is monotonically decreasing with the distance, and the energy provided to the sensors located at $(R,0)$ and $(R,\theta)$ represent local maxima of that provided to the sensors located at the circle edge, therefore we conclude that the solutions $\{(R,\varphi_1), (R,\varphi_2)\}$ correspond to local minima. Once we take into account the contribution of $\mathrm{PB}_3$, the point $(R, \varphi_2)$ is turned into a local maximum, while $(R,\varphi_1)$ remains as minimum. In the general case $|\mathcal{B}| \geq 3$, for $|\mathcal{B}|$ odd, $(R,\varphi_2)$ is a local maximum provided the contribution of the PB located at $(r,\theta/2 + \pi)$, otherwise is local minimum equivalent to $(R,\varphi_1)$. Fig.~\ref{RFatEDGE} summarizes these ideas.
Notice that, there is an alternate pattern of equally-spaced $|\cal B|$ local minima and $|\cal B|$ local maxima at the circle edge. Therefore, we adopt $(R, \varphi_1)$ for finding the value of $r$ that maximizes the minimum average receive power. Then, let us plug this result into \eqref{twoPBs} to get the optimum position $r$ by solving

\begin{figure}[t!]
	\centering
	\includegraphics[width=0.4\textwidth]{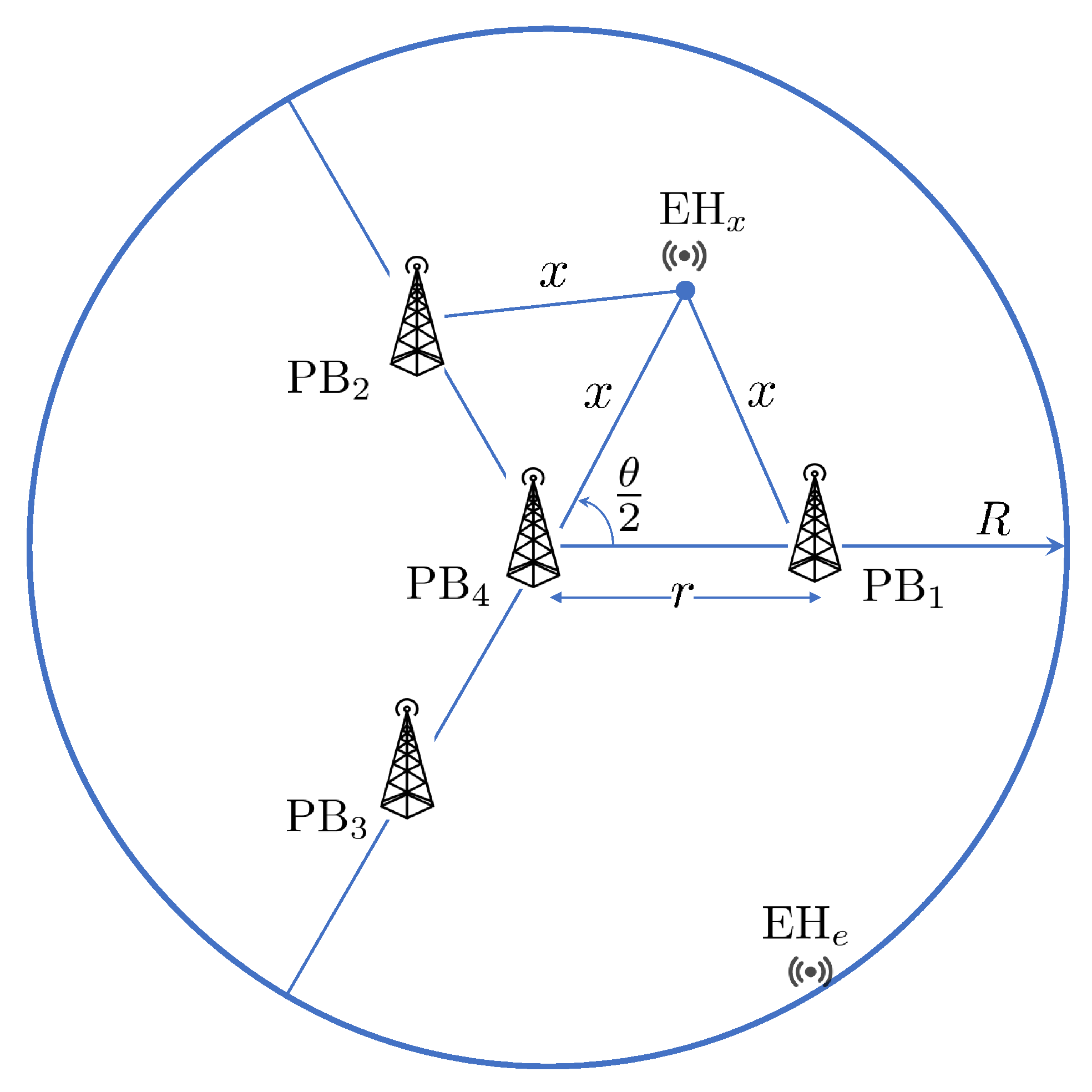}
	\vspace{-3mm}
	\caption{Example of an EC with one PB centered for $|\mathcal{B}| = 4$.}
	\vspace*{-2mm}		
	\label{auxFig2}
\end{figure}

\begin{small}
    \begin{align}\label{derivative}
        \frac{\partial \mathbb{E}[\xi^{RF}_{s}]}{\partial r} = - \frac{2P K (2r-R)}{(r^2 - rR + R^2)^{-\gamma/2-1}} - \frac{P K}{(r + R)^{-\gamma-1}}  = 0. 
    \end{align}
\end{small}
\hspace{-1.5mm}The second fractional term in \eqref{derivative} corresponds to the contribution of $\mathrm{PB}_3$, and can be neglected as the path loss exponent increases, which gives a nearly optimal $r \approx \frac{R}{2}$. Following the same procedure we can arrive to a general analytical approximation $r \approx R \cos{\frac{\theta}{2}}$, which suggests that for the case of $|\mathcal{B}| \in \{1, 2\}$, the placement should be at the center, whereas for $|\mathcal{B}| = 4$ the placement is $r \approx \frac{R \sqrt{2}}{2}$ roughly independent of the propagation conditions. It also guarantees that the minimum contribution will be at the circle's edge rather than at the center. For the general case when $|\mathcal{B}| \geq 5$ is considered, we can use previous results such that $r$ is optimized now for the node with the minimum average received power, which can be either the center or $(R, \varphi_1)$. Unfortunately, the solution of this problem doesn't guarantee the best result for an arbitrary $\cal B$, and solving for the case when the optimum kind of deployment is not known a priori is mathematically intractable. Our proposal is to solve the EC approach algorithmically for two topologies: i) as in Fig.~\ref{auxFig} here called EC; ii) with one PB, the last one, located at the center, thus $\rVert \mathbf{n}_{|\mathcal{B}|}\lVert_2 = 0$, here called EC with one centered PB. Fig.~\ref{auxFig2} depicts the latter topology where $\rVert \mathbf{n}_b \lVert_2 = r$, $\forall b \in \mathcal{B}$, $b \neq |\mathcal{B}|$, and the angular separation between them is $\theta = 2\pi/(|\mathcal{B}| - 1)$.

\begin{figure}[t]
	\centering
	\includegraphics[width=0.45\textwidth]{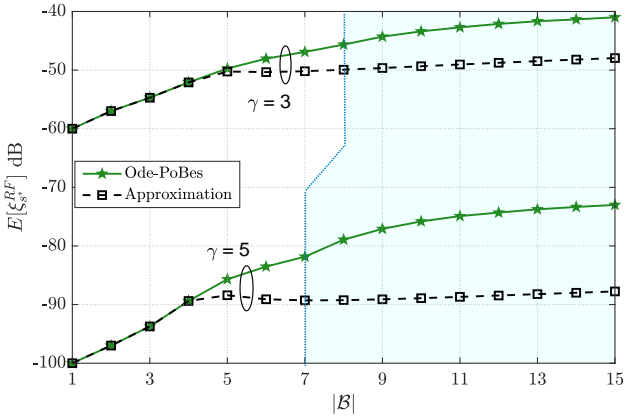}
	\caption{Performance comparison between Ode-PoBes and approximate solution vs $|\mathcal{B}|$ for $\gamma \in \{3, 5\}$. The shaded region denotes the transition to the EC with one PB centered solution using Ode-PoBes solutions.}
	\vspace*{-2mm}		
	\label{approxVSalg1}
\end{figure}

According to Fig.~\ref{auxFig}, the average received power for the node with $s^* = \underset{s}{\operatorname{arg\,min}} \ \mathbb{E}[\xi^\mathrm{RF}_s]$ is
\begin{align}\label{ncentPB}
    \mathbb{E}[\xi^\mathrm{RF}_{s^*}] = \min \Big( \mathbb{E}[\xi^\mathrm{RF}_{c}],\ \mathbb{E}[\xi^\mathrm{RF}_{e}]\Big),
\end{align}
where $\mathbb{E}[\xi^\mathrm{RF}_{e}]$ is the contribution at the worst sensor on the edge with $\varphi = \varphi_1$, thus given by \eqref{twoPBs}, and $\mathbb{E}[\xi^\mathrm{RF}_{c}] = |\mathcal{B}|PKr^{-\gamma}$ is the contribution at the center. Meanwhile, for the deployment in Fig.~\ref{auxFig2}
\begin{align}\label{centPB}
    \mathbb{E}[\xi^\mathrm{RF}_{s^*}] = \min \Big( \mathbb{E}[\xi^\mathrm{RF}_{x}],\ \mathbb{E}[\xi^\mathrm{RF}_{e}]\Big),
\end{align}
where 
\begin{small}
    \begin{equation}
        \mathbb{E}[\xi^\mathrm{RF}_x] \!=\! PK\!\left[\!x^{\!-\!\gamma}\!\!+\!\!\sum_{b=1}^{|\mathcal{B}|-1}\!\left[\! x^2 \!+\! r^2 \!\!-\!\! 2xr \cos{\!\left(\!\theta\!\left(\!b\!-\!\frac{3}{2}\!\right)\!\!\right)\!} \right]^{\!-\!\frac{\gamma}{2}}\!\right]\!,
    \end{equation}
\end{small}
\hspace{-0.9mm}represents the contribution in a sensor equidistant to the center and two adjacent PBs at a distance $x = r/(2\cos{\frac{\theta}{2}})$ and $\varphi = \varphi_1$. Additionally,
\begin{small}
    \begin{equation}
        \mathbb{E}[\xi^\mathrm{RF}_e] \!=\! PK\!\left[\!R^{\!-\!\gamma}\!\!+\!\!\sum_{b=1}^{|\mathcal{B}|-1}\!\left[\! R^2 \!+\! r^2 \!\!-\!\! 2rR \cos{\!\left(\!\theta\!\left(\!b\!-\!\frac{3}{2}\!\right)\!\!\right)\!} \right]^{\!-\!\frac{\gamma}{2}}\!\right]\!,
    \end{equation}
\end{small}
\hspace{-0.9mm}is the average power received at the worst sensor on the edge with $\varphi = \varphi_1$. Finally, the optimal positions correspond to the constellation that maximizes $\mathbb{E}[\xi_{s^*}^\mathrm{RF}]$ against the minimum average contribution when the network has one PB radiating from the circle center, with total power $P_T = |\mathcal{B}|P$, i.e. 
\begin{align}\label{optr}
    r^* = \underset{r}{\operatorname{arg\,max}} \big(\mathbb{E}[\xi^\mathrm{RF}_{s^*}], |\mathcal{B}|PKR^{-\gamma}\big).
\end{align}
Herein, $\mathbb{E}[\xi^{RF}_{s^*}]$ is defined as a parametric function using both $\eqref{ncentPB}$ and $\eqref{centPB}$, thus determined by the chosen topology. The procedure for efficiently determining $r^*$ is detailed in the Optimal DEployment of POwer BEaconS (Ode-PoBes) algorithm\footnote{MatLab implementation of Ode-PoBes algorithm is publicly available at \url{https://github.com/Osmel-dev/Optimization_of_power_beacons}}. Notice that $\Delta r$ denotes the step size of the iterative search.
\begin{algorithm}[t]
\caption{Ode-PoBes}
\begin{algorithmic}[1]\label{alg1}
\STATE \textbf{Input:} $\mathcal{B}, \gamma, R, P, \Delta r$ \label{lin1}
\STATE Set $\xi^* = |\mathcal{B}|PKR^{-\gamma}$
\STATE Set $r^* = 0$
\REPEAT
    \STATE Compute $\mathbb{E}[\xi^\mathrm{RF}_{s^*}]$ using both $\eqref{ncentPB}$ and $\eqref{centPB}$
    \STATE $\theta^* \gets \underset{\theta \in \{\frac{2\pi}{|\mathcal{B}|}, \frac{2\pi}{(|\mathcal{B}|-1)}\}}{\operatorname{arg\,max}} \big(\mathbb{E}[\xi^\mathrm{RF}_{s^*}]\big)$
    \IF{$\xi^* < \max \big(\mathbb{E}[\xi^\mathrm{RF}_{s^*}]\big)$}
        \STATE $\xi^* \gets \max \big(\mathbb{E}[\xi^\mathrm{RF}_{s^*}]\big)$
        \STATE $r^* \gets r$
    \ENDIF
    \STATE $r \gets r + \Delta r$
\UNTIL{$r \geq R$}
\end{algorithmic}
\end{algorithm}
Fig.~\ref{approxVSalg1} depicts a comparison using the results of the Ode-PoBes algorithm and the analytical approximation $\lVert \mathbf{n}_b \rVert_2 \approx R \cos{\frac{\theta}{2}}, \forall b \in \cal B$ for the scenario shown in Fig.~\ref{auxFig}.
In general the approximation holds up to $|\mathcal{B}| = 5$ for $\gamma = 3$, and $|\mathcal{B}| = 4$ for $\gamma = 5$, without an important degradation of the network performance. However, the gap increases with both $|\mathcal{B}|$ and $\gamma$, because this approach aims to benefit more the sensors in the circle's edge at the cost of degrading the contribution at the center. With the increase of $|\mathcal{B}|$, the EC with one centered PB becomes better than EC without PB at the center, having a prompt transition when $\gamma = 5$.

We now provide a worst case complexity analysis for the proposed algorithm. First, for a given $\Delta r$ the required number of iterations is $\lfloor \frac{R}{\Delta r} \rfloor + 1$ and the maximum solution error is $\frac{\Delta r}{2}$. Notice that the argument $\mathbb{E}[\xi_e^\mathrm{RF}]$ determines the computational cost of $(10)$, and $(6)$ proportionally scales up with $|\mathcal{B}|$. Similarly, each argument in $(11)$ has a cost proportional to $|\mathcal{B}|$. Then, the most costly step in Ode-PoBes, step $6$, requires $\mathcal{O}(|\mathcal{B}|)$ operations, which in turn constitutes the computational cost of each iteration of the Ode-PoBes algorithm.

\subsection{Interior-point method approach (IPM)}
Interior-point methods, for solving inequality-constrained optimization problems rely on gradient-based equations, therefore require a differentiable objective function \cite{boyd2004convex}. Objective function in \eqref{P2a} is neither smooth nor differentiable, therefore we resort to generalized mean approximation which can be used as a smooth estimate of the $\min$ function \cite{bullen2013handbook}:
\begin{align}\label{powermeanapp}
\min_{s} \mathbb{E}[\xi^{RF}_s] \approx f_k(\bm{\xi}) = \Big(\frac{1}{|\cal S|} \sum_{s=1}^{|\cal S|} \xi_s^{k}\Big)^\frac{1}{k},
\end{align} 
%
where $\bm{\xi}$ represents a $1 \times |\cal S|$ vector comprising the average power received at each $s$, thus $\xi_s = \mathbb{E}[\xi^{RF}_s]$. The previous relation tends to be an equality as $k \to -\infty$ but at the expense of an increasing computational cost. Other differentiable approximations are addressed in \cite{lange2014applications} for gradient-based learning algorithms, using the $\mathrm{softmax}$, $\min_{s} \mathbb{E}[\xi^{RF}_s] \approx \frac{\sum_{s=1}^{|\cal S|} \xi_s e^{k \xi_s}}{\sum_{s=1}^{|\cal S|} e^{k \xi_s}}$ \cite[eq.(7)]{lange2014applications}, and the $\mathrm{quasimax}$, $\min_{s} \mathbb{E}[\xi^{RF}_s] \approx \frac{1}{k} \log \Big(\sum_{s=1}^{|\cal S|} e^{k \xi_s}\Big)$ \cite[eq.(8)]{lange2014applications},	 functions. Here $k$ is a real constant that determines whether the approximation tends to the minimum function ($k < 0$) or to the maximum one ($k > 0$). All these approximations perform alike, but since the gradient of the objective must be provided to the IPM algorithm, we adopt \eqref{powermeanapp} whose derivative is more tractable. We can now transform \textbf{P2} to solve it using IPM based on the logarithmic barrier function as
\begin{subequations}\label{P2.1}
    \begin{alignat}{2}
    \mathbf{P2.1:} \quad& \underset{\{\mathbf{n}_b\}, \mathbf{t}}{\mathrm{minimize}} && -f_k(\bm{\xi}) - \mu \sum_{b=1}^{|\cal B|} \ln(t_b), \\
    &\text{subject to} &&\quad \lVert \mathbf{n}_b \rVert_2 - R + t_b = 0,\ \forall b \in \cal B,
    \end{alignat}
\end{subequations}
where $t_b > 0$ is the $b^\text{th}$ component of the slack variable $\mathbf{t} \in \mathbb{R}^{|\mathcal{B}| \times 1}$ for turning the set of $|\mathcal{B}|$ inequalities in \textbf{P1} into equality constraints. Moreover, $\mu > 0$ is the barrier parameter that must be chosen small enough to improve the accuracy of the approximation. The next step involves the solution of the system of equations derived from the \textit{Karush-Kuhn-Tucker} (KKT) conditions \cite{boyd2004convex}
\begin{align}
    \nabla \mathcal{L}(\{\mathbf{n}_b\}, \mathbf{t}, \bm{\lambda}) &= 0,\label{lagran}\quad \forall b \in \cal B,\\
    \lVert \mathbf{n}_b \rVert^2 - R + t_b &= 0,\quad \forall b \in \cal B,
\end{align}
where each entry of $\bm{\lambda} = \{ \lambda_b \}$ corresponds to the Lagrange multiplier associated with the $b^\text{th}$ equality constraint. The partial derivatives of the Lagrangian in $\eqref{lagran}$ are given by
\begin{align}
 \frac{\partial \mathcal{L}}{\partial \mathbf{n}_b} &= f_k(\bm{\xi})^{1-k} \frac{1}{|\cal S|}\sum_{s = 1}^{|\cal S|} \xi^{k-1}_s\frac{\partial \xi_s}{\partial \mathbf{n}_b} + 2 \lambda_b \mathbf{n}_b,\label{Partial:1}\\
\frac{\partial \mathcal{L}}{\partial t_b} &= \frac{\mu}{t_b} + \lambda_b,\label{Partial:2}
\end{align}
where
\begin{align}
\frac{\partial \xi_s}{\partial \mathbf{n}_b} &= \gamma P K \lVert \mathbf{u}_s - \mathbf{n}_b\rVert^{-\gamma - 2}_2(\mathbf{u}_s - \mathbf{n}_b) .
\end{align}

In general, the system (or the problem) is solved through a sequence of linear equality constrained quadratic problem applying the Newton's method \cite{5464889}, or the conjugate gradient method \cite{doi:10.1137/0720042}. The performance of the IPM vs $k$ is evaluated in Table~\ref{tab:IPMs} for $|\mathcal{B}| \in \{3, 9, 15\}$ using normalized values. For instance, we normalize the elements of each triplet with respect to $\underset{k}{\max}\ \mathbb{E}[\xi^{RF}_{s^*}]$ for given $\gamma$ and $|\cal B|$. 
\begin{table}[t]
    \centering
    \begin{threeparttable}[b]
        \caption{Normalized values\tnote{} of $\mathbb{E}[\xi^{RF}_{s^*}]$ vs $k$ for $\gamma \in \{3, 5\}$ and $R=100$m.} 
        \label{tab:IPMs}
        \begin{tabular}{c c c}
            \thickhline
                $k$ & $\gamma = 3$ & $\gamma = 5$ \\
            \hline
                $-5$ & $\{0.9851,\ 0.9101,\ 0.8989 \}$ & $\{0.9921,\ 0.8899,\ 0.8817\}$ \\
                $-10$ & $\{0.9944,\ 0.9525,\ 0.9037\}$ & $\{0.9973,\ 0.9536,\ 0.9768\}$ \\
                $-15$ & $\{0.9975,\ 0.9740,\ 0.9726\}$ & $\{0.9994,\ 0.9785,\ 0.9703\}$ \\
                $-20$ & $\{0.9990,\ 0.9873,\ 0.9857\}$ & $\{0.9996,\ 0.9919,\ 1.0000\}$ \\
                $-25$ & $\{0.9995,\ 0.9947,\ 0.9941\}$ & $\{1.0000,\ 1.0000,\ 0.9814\}$ \\
                $-30$ & $\{1.000,\ 1.000,\ 1.000\}$ & Not feasible \\
            \thickhline
        \end{tabular}
        \begin{tablenotes}
            \item[] The solutions in the triplets are ordered according to $|\mathcal{B}| \in \{3, 9, 15\}$
        \end{tablenotes}
    \end{threeparttable}    
\end{table}
As $k$ decreases, the approximation \eqref{powermeanapp} converges in almost all cases; however, it also requires more computational effort, but still for small $|k|$, the results are quite good. Notice that for $k=-30$ IPM becomes unstable for large values of the path-loss exponent, therefore compromising the reliability of the solution. Hereafter, we adopt $k = -25$ since it gives the best results in terms of stability and accuracy.
\subsection{Evolutionary computation algorithms}
Modern meta-heuristics methods such as \textit{genetic algorithms} (GA) and \textit{particle swarm optimization} (PSO) are stochastic approaches, thus, suitable for dealing with optimization problems where the objective function is highly nonlinear and nondifferentiable. GA is inspired by the process of evolutionary biology, in which subsequent generations evolve from previous throughout selection, crossover and mutation \cite{yang2018optimization}. At each step, a group of individuals of the current population are selected for the next generation according to their performance against the objective function. Each of them is a possible solution of the optimization problem at a given iteration. They might also be used for crossover and mutation which helps to search over the feasible solution set avoiding to get trapped at a local optimum. This allows us to solve directly the problem $\mathbf{P2}$, without the approximation in \eqref{powermeanapp}.

Swarm intelligence-based algorithms such as PSO use the principles of self-organization of the so-called agent particles \cite{yang2018optimization}. Each particle represents a candidate solution of the optimization problem. PSO aims to find the global minimum by moving the agents in a quasi-static manner within the domain of an objective function, starting from initial positions and velocity. Then, at $i^\mathrm{th}$ iteration the position of the $b^\text{th}$ PB in the $q \in \mathcal{Q}$ particle is $\{\mathbf{z}_{q,i,b} \in \mathbb{R}^{2 \times 1}|\forall b \in \cal B\}$.
Therefore, each particle's position set is made up of a candidate solution in $\mathbf{P2}$. Let us define the objective function as

\begin{small}
    \begin{align}\label{fitnessFunct}
    	f_\mathrm{PSO}\!(\!\{ \mathbf{z}_{q,i,b}|\forall b \in \mathcal{B} \}\!)\! \!=\! \underset{s}{\min} \ \mathbb{E}[\xi^{RF}_s] \!-\! \epsilon \sum_{b=1}^{|\mathcal{B}|} \frac{1}{\big(\lVert \mathbf{z}_{q,i,b} \rVert_2 \!-\! R\big)},
    \end{align}
\end{small}
\hspace{-1.5mm}where $\mathbb{E}[\xi^{RF}_s]$ is computed according with \eqref{objFunc} but replacing $\mathbf{n}_b$ for $\mathbf{z}_{q,i,b}$, and the second term is a barrier function whose impact on the problem solution is limited by making $\epsilon\rightarrow 0$. At the iteration $i'$, the particles update their position and velocity considering its initial values, local best position $\mathrm{pbest}(q,i')$, and the global best position $\mathrm{gbest}(q)$ \cite{5518452}, where
\begin{align*}
    \mathrm{pbest}(q,i') &= \underset{i = 1, 2, \ldots, i'}{\operatorname{arg\,min}} \ f_\mathrm{PSO}(\{ \mathbf{z}_{q,i,b}|\forall b \in \mathcal{B} \}), \ \forall q \in \cal Q \\
    \mathrm{gbest}(q) &= \underset{\substack{i = 1, 2, \ldots, i' \\ \forall q \in \cal Q}}{\operatorname{arg\,min}} f_\mathrm{PSO}(\{ \mathbf{z}_{q,i,b}|\forall b \in \mathcal{B} \}),
\end{align*}
to find the best of all local solutions until the objective no longer improves.
Once \textbf{P2} is solved, one needs to minimize the number of deployed PBs that satisfies the energy outage probability requirement in \eqref{P1b}. Algorithm~\ref{alg2} shows the iterative procedure for finding the minimum $|\mathcal{B}|$.
\begin{algorithm}[t]
\caption{Computation of the minimum $|\mathcal{B}|$}
\begin{algorithmic}[1]\label{alg2}
\STATE \textbf{Input:} $\gamma, R, P_T, \Delta r, \xi_0, \zeta $ \label{alg2:1}
\STATE Set $|\mathcal{B}| = 1$ \label{alg2:2}
\WHILE{$\mathbb{P}[\xi_s^\mathrm{RF} \leq \xi_0] > \zeta$}
\STATE $|\mathcal{B}| \gets |\mathcal{B}| + 1$ 
\STATE $P \gets \frac{P_T}{|\mathcal{B}|}$
\STATE Solve \textbf{P2}
\ENDWHILE
\end{algorithmic}
\end{algorithm}
At each iteration, \textbf{P2} is solved until the energy outage condition is guaranteed. Remember that the distribution of $\xi_s^\mathrm{RF}$ obeys \eqref{gCHI}. In the next section we present optimal deployment results under different network conditions. 

\subsection{Practical implementation considerations}
In practice, the performance of the algorithm for finding the PBs' positions depends on how well the path loss model fits the actual conditions. For instance, recent measurement campaigns in wireless sensor networks corroborate the dependence of the path loss exponent on the environment's characteristics \cite{8909137}. Hence, different sub-regions of the network might require different path loss models. Moreover, the inherent properties of IoT networks (e.g. small antenna heights, low transmission power, and stationary nodes) limit the applicability of traditional propagation models \cite{7809115}. Similarly, the proposed algorithm framework guarantees meeting the energy outage requirement provided the channel fading distribution is accurately known beforehand.

Therefore, opportunistically selecting appropriate channel models, at both large (path-loss) and small scale (fading distribution), is essential. Machine learning methods are potential candidates for such a task, that must also provide confidence when a large set measurements is available \cite{aldossari_machine_2019}. The risk of channel modeling/prediction errors must be taken into account in the optimization framework as well. Finally, we can utilize the solutions from Ode-PoBes as a good initial guess for any of the metaheuristic algorithms before treated, which can improve the performance of the final deployment.

\section{Numerical Results}\label{numerical}
In this section, we present numerical results on the optimal PBs deployment to meet a certain energy outage constraint. Additionally, we provide insights on the maximum coverage area and the impact of multi-antenna schemes for massive WET. Unless we state the contrary, the simulations are based on the parameters listed in Table~\ref{tab:param}.
\begin{table}[t]
    \centering
    \caption{Simulation Parameters}
    \label{tab:param}
    \begin{tabular}{c c | c c}
        \thickhline
            \textbf{Parameter} & \textbf{Value(s)} & \textbf{Parameter} & \textbf{Value(s)}\\
        \hline
            $R$ & $100$~m & $|\mathcal{S}|$ & $1000$ \\
            $P_T$ & $10$~W & $\gamma$ & $3$ \\
            $\xi_0$ & $-22$~dBm \cite{8742574} & $\kappa$ & $3$ \\
            $K$ & $1$ & \\
        \thickhline
    \end{tabular}
\end{table} 
\subsection{On the optimal deployment of PBs}
\begin{figure}[t]
	\centering
	\includegraphics[width=0.5\textwidth]{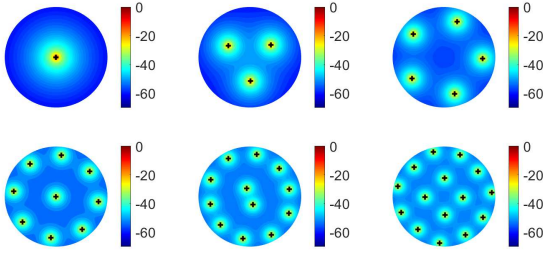}
	\vspace*{-5mm}
	\caption{Heat map of the available average power in the circle area for different number of PBs and $P_T = 1$. The black crosses represent the optimal PB solutions using IPM.}
	\label{HeatMap}
\end{figure}

\begin{figure}[t!]
	\centering
	\subfigure{\includegraphics[width=0.45\textwidth]{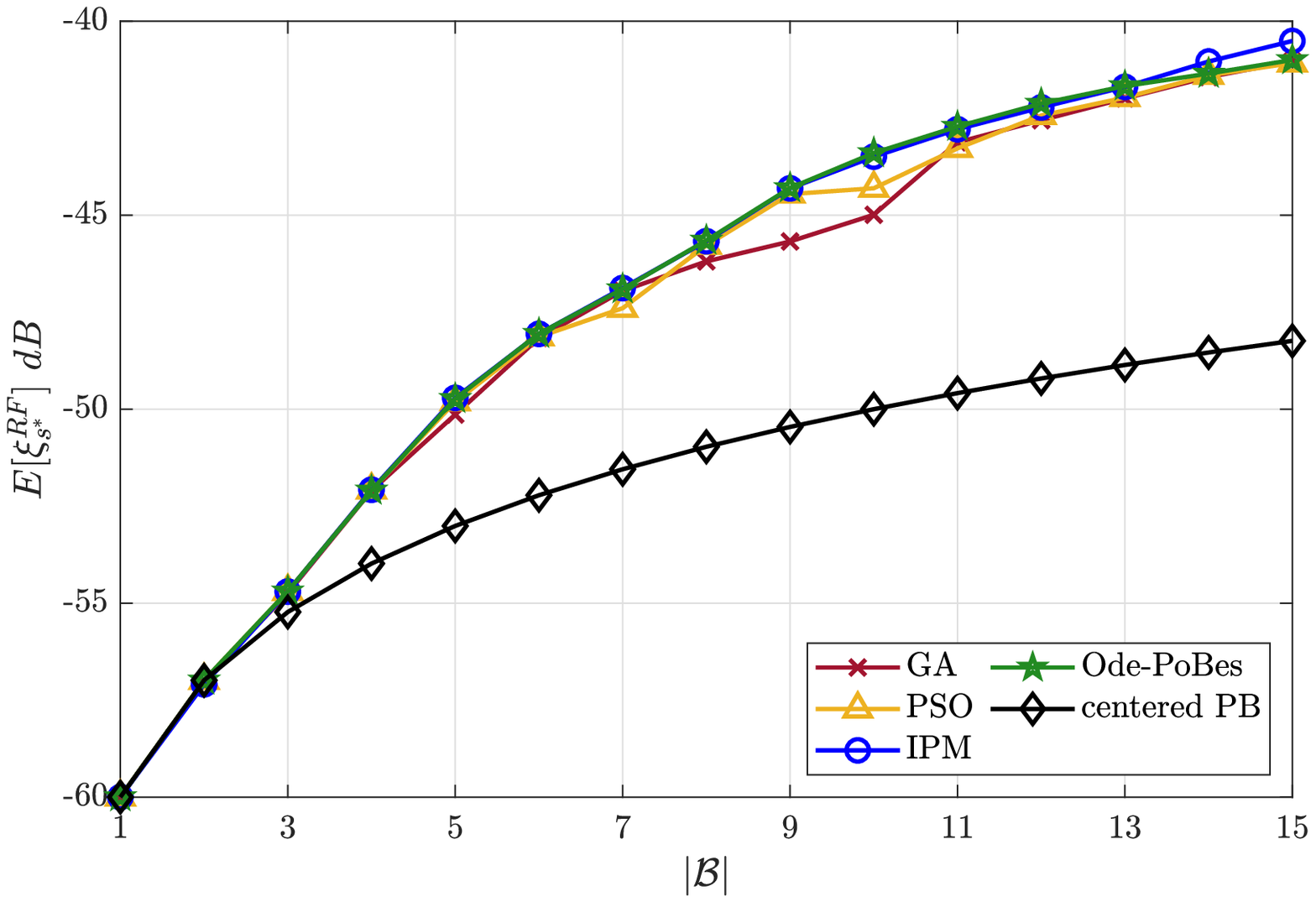}} \\
	\subfigure{\includegraphics[width=0.45\textwidth]{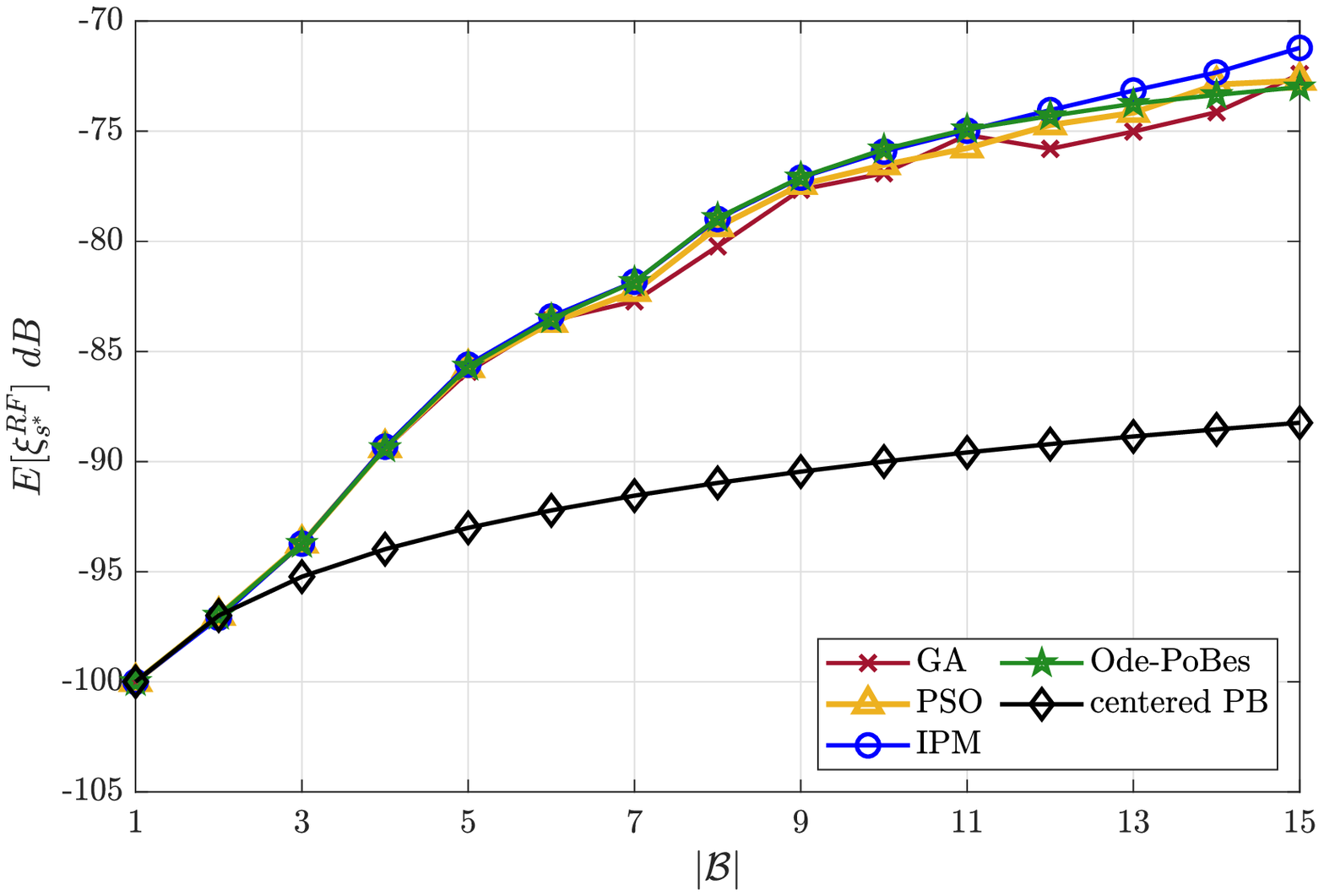}}
	\caption{Available average power at the worst position for optimal PB placement, $P=1$, and: $\gamma = 3$(top), $\gamma = 5$(bottom)}
	\label{optimalPlacement}
\end{figure}

The optimal positions after solving $\mathbf{P2}$ are represented in Fig.~\ref{HeatMap} for different number of PBs; while we show the corresponding average power along the circle area as a heat map.
The optimal PBs' positions are equidistant from the circle' center, and form a symmetric distributed pattern with more rings as either the number of PB increases or the propagation conditions worsen. As we include more PBs in the network, the performance of the worst sensor improves in terms of average received power $\mathbb{E}[\xi^{RF}_{s^*}]$ under the different optimization approaches, which can be also observed in Fig.~\ref{optimalPlacement}. Here, as a benchmark, we also present the case of a centered PB radiating with total power $P_T$, which is equivalent to place all $|\mathcal{B}|$ PBs at the center.
In general all methods outperform the benchmark, while the performance gap increases with $|\mathcal{B}|$. The special case $|\mathcal{B}| = 2$ is equivalent to place a single PB at the circle' center as all methods agree, although doubling the power. Ode-PoBes and IPM stand out over the others in terms of stability in the final solution. The optimization is limited to $15$ PBs due to the poor convergence of GA compared with the IPM approach. In fact, the classical GA approach doesn't converge towards the desired solution as the complexity of the problem increases. A more stable outcome is obtained with PSO, as compared with IPM as a reference.

\begin{figure}[t!]
	\centering
	\includegraphics[width=0.45\textwidth]{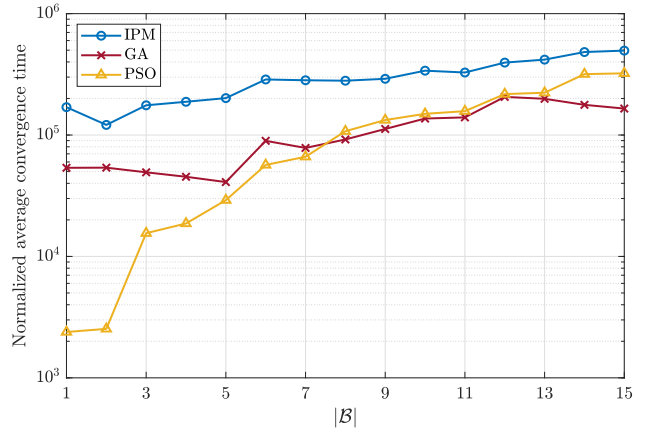}
	\caption{Normalized average convergence time vs $|\mathcal{B}|$.}
	\label{time}
\end{figure}

The results obtained with Ode-PoBes follow the trend of the ones with the IPM, but the computational cost is smaller for the former. Indeed, Ode-PoBes searches for the optimum positioning without requiring derivative computations. Fig.~\ref{time} shows the normalized average convergence time with respect to the time required for Ode-PoBes.
Notice that the Ode-PoBes algorithm is superior by four order of magnitude since it reduces the dimension of the optimization variable. Although the other methods presented searches for individual positions there is a gain in the stochastic approaches over the IPM, given they do not require to compute the gradient at each iteration. However, the gap is reduced as $|\mathcal{B}|$ increases, i.e. as the number of possible constellations increases. Finally, for $|\mathcal{B}| \geq 7$ the evolutionary algorithms approaches exhibit a close performance. 

\subsection{On the solutions of \textbf{P1}}
\begin{figure}[t!]
	\centering
	\includegraphics[width=0.45\textwidth]{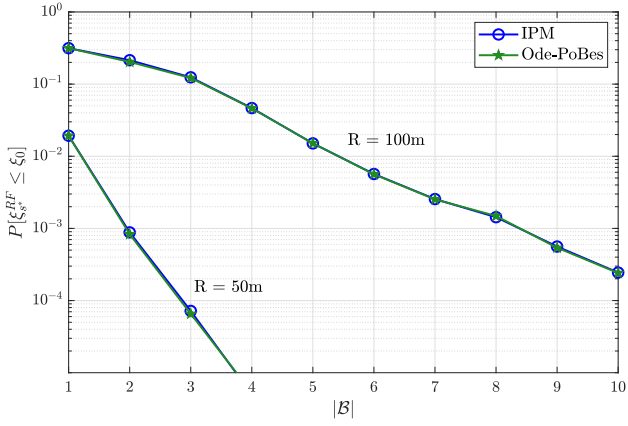}
	\caption{Energy outage probability vs $|\mathcal{B}|$, for $R \in \{50, 100\}$~m.}
	\label{eOutage1}
\end{figure}

Let us consider the scenario where wireless charging is provided as a service to the IoT devices, and each PB is powered by an external wired source. In order to obtain profits, the service provider chooses to distribute a fixed power budget of $P_T = 10$~W among the individual PBs according to \eqref{P1c}. Fig.~\ref{eOutage1} depicts the energy outage probability as a function of $|\mathcal{B}|$, solving the optimum positions with IPM and Ode-PoBes. Hereinafter, IPM will be our benchmark strategy provided its accuracy over GA and PSO. The monotonic decrease of the energy outage probability for both approaches, proves the effectiveness of the distributed PBs approach when improving the network reliability. 

\begin{figure}[t!]
	\centering
	\subfigure{\includegraphics[width=0.45\textwidth]{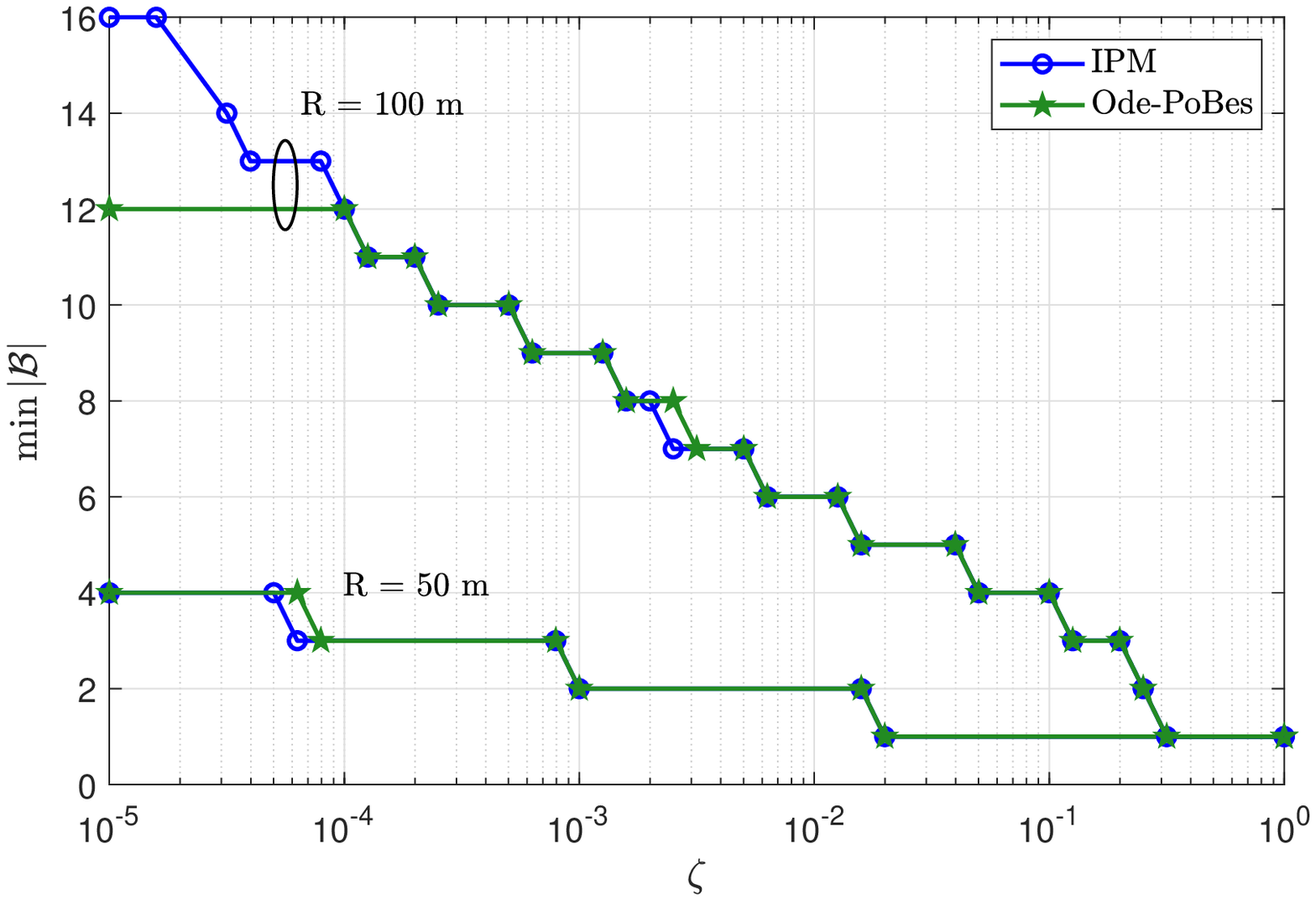}}
	\\ 
	\subfigure{\includegraphics[width=0.45\textwidth]{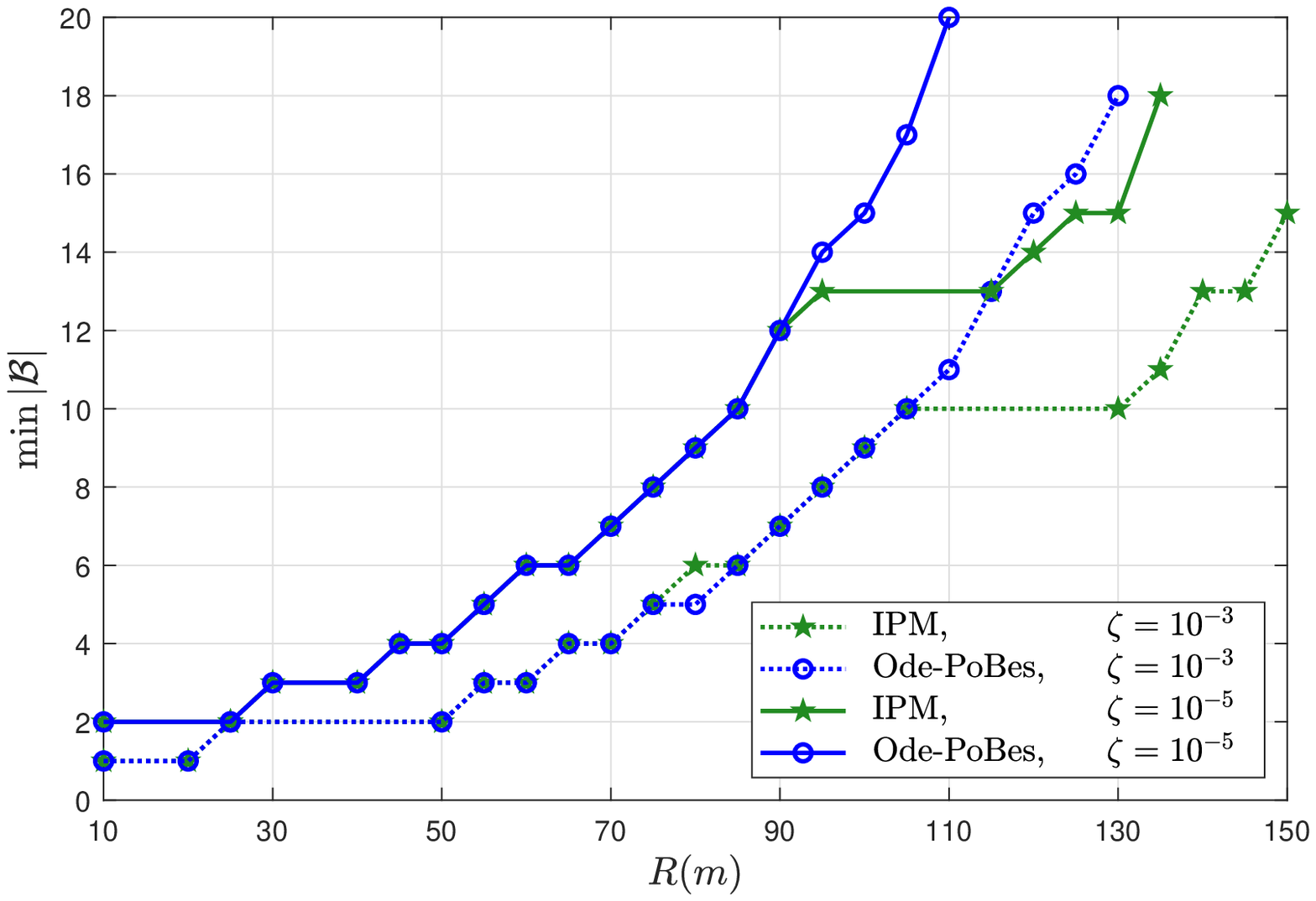}}
	\caption{Solutions of $\mathbf{P1}$ using Ode-PoBes and IPM: i) $\min |\mathcal{B}|$ vs network reliability $\zeta$ for $R \in \{50, 100\}$~m (top); ii) $\min |\mathcal{B}|$ vs $R$ for $\zeta \in \{10^{-3}, 10^{-5}\}$ (bottom).}
	\label{Nmin}
\end{figure}

For this setup, we also present the solutions of $\mathbf{P1}$ in Fig.~\ref{Nmin} as a function of $\zeta$ and $R$. Fig.~\ref{Nmin} (top) shows that tightening the QoS agreements requires more PBs to be deployed; whereas Fig.~\ref{Nmin} (bottom) depicts a stable increment in the number of PBs needed to meet the target QoS constraint as the network gets larger. In general, the fluctuations in the minimum number of PBs ($\min |\mathcal{B}|$) reflect the dominant impact of the distance-dependent loss. Notice that Ode-PoBes outperform IPM's solutions when serving larger areas with tight QoS requirements. Particularly, Ode-PoBes finds an optimal deployment with $4$ PBs less than IPM when $\zeta = 10^{-5}$ and $R = 100$~m. Moreover, Ode-PoBes makes possible to cover wider areas as compared to the IPM based solutions with the same QoS requirements. The gap between Ode-PoBes and IPM is due to numerical approximations that affect the symmetry of the final solution. In fact, IPM solves $|\mathcal{B}|$ individual positions within the space of possible constellations, which doesn't guarantee symmetric layers with respect to the origin. Meanwhile, Ode-PoBes finds $(r^*, \theta^*)$ that maximizes the $\mathbb{E}[\xi_{s^*}^\mathrm{RF}]$ using two predefined symmetric constellations. 
\subsection{On the maximum coverage area}
\begin{figure}[t!]
	\centering
	\subfigure{\includegraphics[width=0.45\textwidth]{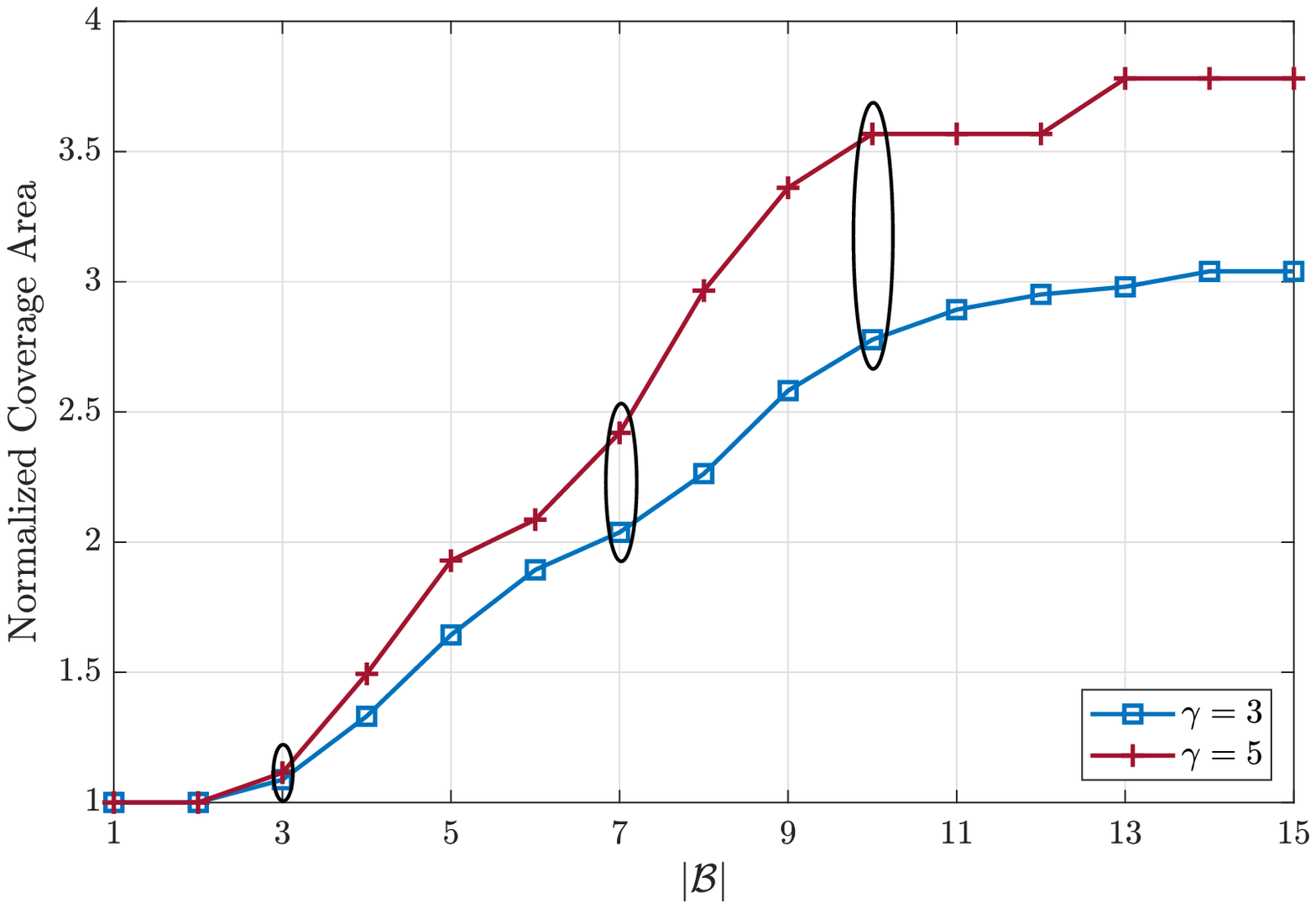}}
	\subfigure{\includegraphics[width=0.45\textwidth]{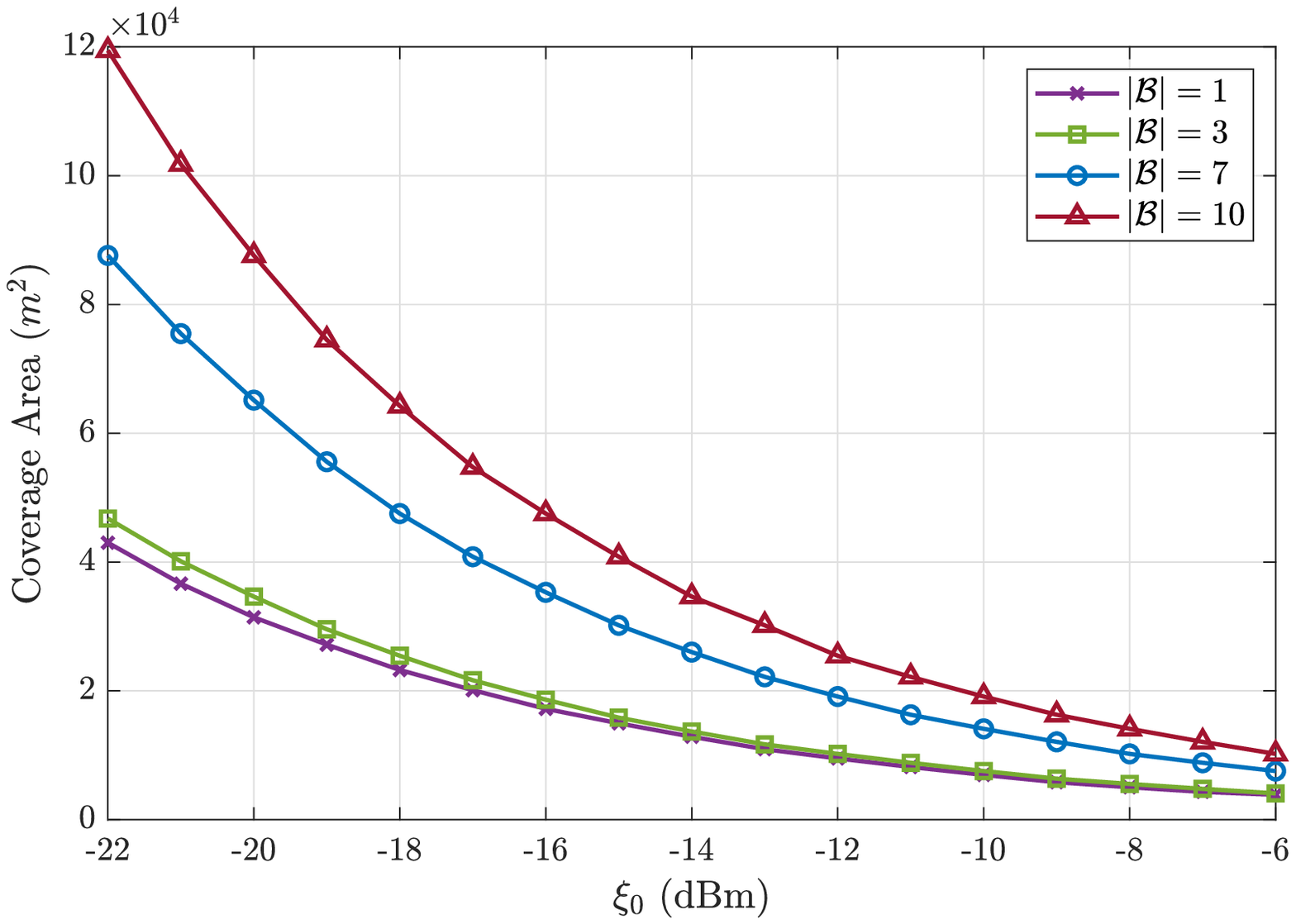}}
	\caption{Maximum coverage area results: i) Normalized to the maximum coverage area with a centered PB vs $|\mathcal{B}|$ for $\gamma \in \{3, 5\}$ (top); ii) Coverage area vs $\xi_0$ for $|\mathcal{B}| \in \{1, 3, 7, 10\}$ (bottom).}
	\label{coverageArea}
\end{figure}

In Fig.~\ref{coverageArea}, we show the maximum coverage area results with respect to $\mathbb{E}[\xi^{RF}_{s^*}]$ using the minimum average incident RF power obtained with Ode-PoBes. Clearly, the distributed deployment outperforms the centered PB approach, and the performance gap increases with the number of PBs and the path-loss exponent. Our proposal shows that bringing the PBs closer to the worst-positioned sensors is more effective to overcome the distance-dependent loss than increasing the power of a single PB. Additionally, as the sensitivity $\xi_0$ of the harvesting circuitry increases, not just the coverage area but the gap among different configurations diminish. In fact, as the power budget is shared by more PBs, the received power at reference distance of $1$~m decreases by a factor of $|\cal B|$ with respect to the centered PB approach. Moreover, it allows to extend the coverage area without increasing the level of RF-EMF in the proximity of the PBs. Notice that in Fig.~\ref{coverageArea} (top) the points $|\mathcal{B}| \in \{3, 7, 10\}$ have been highlighted, so that the readers can observe the correspondence with the results in Fig.~\ref{coverageArea} (bottom) when $\xi_0 = -22$~dBm.
\subsection{On CSI-free multi-antenna WET}
Performance of multiple-antenna strategies for powering a massive number of energy efficient devices are analyzed in \cite{8760520} using one antenna (OA), all antennas at once (AA) and switching antennas (SA) schemes. We only focus here on the SA scheme given that OA can be seen as an special case when only a single antenna is used. On the other hand, the performance claimed for AA in \cite{8760520} is valid only for equal mean phases along the transmit antennas, which is difficult to hold in practice. Under SA strategy, each PB transmits with full power by one antenna at the time, in a way that the whole array is utilized within a channel coherence block. Additionally, SA preserves the harvested energy as in the case of single-antenna PBs (although boosting the energy diversity), harvested while the variance is a function of the spatial correlation among the antennas.

For this setup, let us consider that each PB is equipped with multiple antennas denoted by the set $\mathcal{A}$. We keep the assumption on the fading distribution for each antenna at PBs to transmit independent energy-carrying signals.
\begin{figure}[t!]
	\centering
	\includegraphics[width=0.5\textwidth]{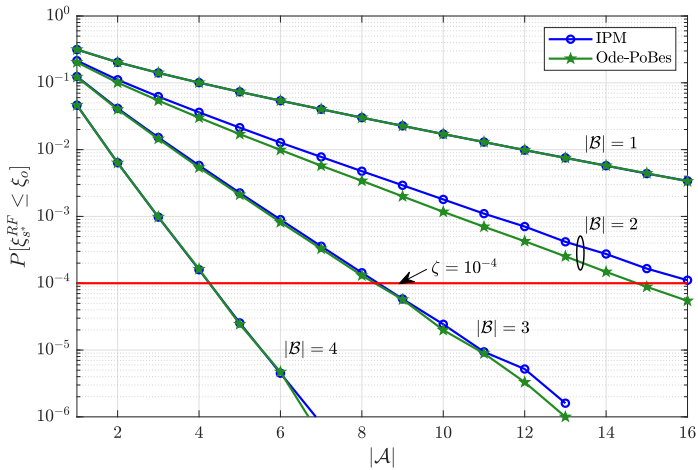}
	\caption{Energy outage probability versus number of antennas, for $|\mathcal{B}| \in \{1,2,3,4\}$.}
	\label{energyOutage}
\end{figure}
The use of SA scheme, as Fig.~\ref{energyOutage} depicts, improves the system performance in terms of reducing the outage probability at the worst sensor as the number of antennas increases. As an example, we constraint the system to $\zeta = 10^{-4}$, and it can be observed that the number of PBs have a greater impact on the system performance than the number of antennas. In fact, the outage probability does not improve significantly as the number of antennas increases when $|\mathcal{B}| = 1$, but once more PBs are distributed in the network, the performance impact of the number of antennas grows significantly. The readers can observe that the deployment of an additional PB could roughly be compensated by doubling the number of antennas at the existing PBs.

\section{Conclusion}\label{conclusions}
In this paper, we studied the mixed-integer programming problem of PBs' deployment optimization with aim to wirelessly charge a massive number of IoT devices under energy outage constraints. We first found the optimal PBs' positions by using a heuristic method based on equally-far-from-center restrictions, and then we solve it algorithmically. As a benchmark, we compute the optimal positioning problem using interior-point methods by approximating the objective with the generalized mean, and also through evolutionary algorithms, without prior knowledge about devices' locations. Numerical results show that our algorithm performs similar to the benchmark approaches, but it converges much faster given the constraints imposed to the PBs' positions. As compared with the centered PB approach, properly distributed PB deployments introduce improvements in terms of energy outage probability and coverage area. Additionally, it was shown the dominant impact of the distance-dependent loss on the solution of the minimization of the number of required PBs under a certain outage constraint as the radius increases. Finally, we presented energy outage results using CSI-free WET schemes for the case of multiple antennas at the PBs. Although having multi-antenna PBs always introduces improvements, we found that the number of deployed PBs impacts stronger the system performance. Indeed, considering the same total power, the performance obtained when doubling the number of antennas at the PBs can be alternatively attained by properly deploying an additional PB which reduces the hardware complexity. Our results provide valuable insights for designing practical WET setups by answering how many PBs are needed, and their corresponding locations, for powering certain area with average power or energy outage QoS constraints.

An attractive future work is to study the impact of knowing the position of the sensors in the PBs' deployment optimization. Notice that the deployment's symmetry assumption is crucial in this work. Deviations from the optimal deployment caused by obstacles or forbidden installation places can significantly degrade the incident power at the worst position. Hence, we could consider the possibility of having moving PBs in order to provide flexibility to network changes. Different from previous works, we can optimize the PBs' positions with the help of probabilistic machine learning tools, which provide low-complexity solutions to large problems. 

\bibliographystyle{IEEEtran}
\bibliography{IEEEabrv,references}
\end{document}